# On the Mean-Square Performance of the Constrained LMS Algorithm

Reza Arablouei, Kutluyıl Doğançay, and Stefan Werner

*Abstract*—The so-called constrained least mean-square algorithm is one of the most commonly used linear-equality-constrained adaptive filtering algorithms. Its main advantages are adaptability and relative simplicity. In order to gain analytical insights into the performance of this algorithm, we examine its mean-square performance and derive theoretical expressions for its transient and steady-state mean-square deviation. Our methodology is inspired by the principle of energy conservation in adaptive filters. Simulation results corroborate the accuracy of the derived formula.

## I. Introduction

CONSTRAINED adaptive filtering algorithms are powerful tools tailored for applications where a parameter vector need be estimated subject to a set of linear equality constraints. Examples of such applications are antenna array processing, spectral analysis, linear-phase system identification, and blind multiuser detection. The deterministic constraints are usually construed from some prior knowledge about the considered problem such as directions of arrival in antenna array processing, linear phase in system identification, and spreading codes in multiuser detection. In some other applications, specific linear equality constraints can help improve robustness of the estimates or obviate a training phase [1]-[3].

The constrained least mean-square (CLMS) algorithm proposed in [4], [5] is a popular linearly-equality-constrained adaptive filtering algorithm. It was originally developed for array processing as an online linearly-constrained minimum-variance (LCMV) filter [2]. The CLMS algorithm implements stochastic gradient-descent optimization. Hence, it is relatively simple in structure and computational complexity. It is also capable of adapting to slow changes in the system parameters or the statistical properties of the input data. It has been widely utilized in applications pertaining to adaptive LCMV filtering, particularly adaptive beamforming [6]-[12]. Several other linearly-constrained adaptive filtering algorithms have been proposed, which are computationally more demanding compared with the CLMS algorithm but offer improved convergence speed or steady-state performance [13]-[22].

Performance analysis of the constrained adaptive filtering algorithms is often challenging since the incorporation of the constraints makes their update equations more complex than those of the unconstrained algorithms. In [4], mean performance of the CLMS algorithm is analyzed. It is shown that, for an appropriately selected step-size, the CLMS algorithm converges to the optimal solution in the mean sense, i.e., the CLMS algorithm is asymptotically unbiased. Moreover, using the analysis results from [23]-[25], a stable operating range for the step-size as well as lower and upper bounds for the steady-state misadjustment of the CLMS algorithm are specified. These bounds are derived under the assumption that the input vectors are temporally independent and have multivariate Gaussian distribution. In [6] and [7], the mean-square performance of the CLMS algorithm is analyzed and its theoretical steady-state mean output-power and misadjustment are computed. The former studies the behavior of the weight covariance matrix and the latter considers the weight-error covariance matrix. However, the analyses in [6] and [7] are carried out for the particular application of adaptive beamforming where the objective is to minimize the filter output energy and there is no observed reference or training signal. Moreover, the analytical methods employed in these works are not suitable for studying the dynamics of the algorithm's mean-square deviation (MSD). MSD is the expectation of the squared norm of the difference between the estimate vector and the optimal solution vector. It a particularly important representative of performance when the objective is primarily to identify the unobserved parameters of an underlying system that governs a linear relation between the input and output of the system while the parameter estimates are required to satisfy certain linear equality constraints. The examples of such applications abound in estimation and control theories [1], [26]-[28].

In this letter, we take a fresh look into the mean-square performance of the general-form CLMS algorithm from the perspective of a technique based on the energy conservation arguments [29]. We study the mean-square convergence of the CLMS algorithm and find the stable operating range for its step-size parameter. Then, we derive theoretical expressions for the transient as well as steady-state values of the MSD of the CLMS algorithm. Following the same line of analysis, we also derive a theoretical expression for the steady-state

R. Arablouei and K. Doğançay are with the School of Engineering and the Institute for Telecommunications Research, University of South Australia, Mawson Lakes SA 5095, Australia (email: arary003@mymail.unisa.edu.au, kutluyil.dogancay@unisa.edu.au).

S. Werner is with the Department of Signal Processing and Acoustics, School of Electrical Engineering, Aalto University, Espoo, Finland (email: stefan.werner@aalto.fi).

misadjustment of the CLMS algorithm and show that it is in agreement with the one given in [6] and [7]. Our simulation results exhibit a good agreement between the theoretically predicted and experimentally found values of the MSD. Therefore, the presented analysis sheds valuable light on the mean-square performance of the CLMS algorithm.

## II. ALGORITHM

Consider a linear system where, at each time instant $n \in \mathbb{N}$, an input vector $\mathbf{x}_n \in \mathbb{R}^{L \times 1}$ and an output scalar $y_n \in \mathbb{R}$ are related via

$$y_n = \mathbf{x}_n^T \mathbf{h} + v_n. \quad (1)$$

Here, $\mathbf{h} \in \mathbb{R}^{L \times 1}$ is the system parameter vector, $v_n \in \mathbb{R}$ is the background noise, and $L \in \mathbb{N}$ is the order of the system. An adaptive filter of order $L$, with tap-coefficients vector $\mathbf{w}_n \in \mathbb{R}^{L \times 1}$, is employed to find an estimate of $\mathbf{h}$ from the observed input-output data. In addition, at every iteration, $1 \leq K < L$ linear equality constraints are imposed upon $\mathbf{w}_n$ such that to have

$$\mathbf{C}^\top \mathbf{w}_n = \mathbf{f} \quad (2)$$

where $\mathbf{C} \in \mathbb{R}^{L \times K}$ and $\mathbf{f} \in \mathbb{R}^{K \times 1}$ are the constraint parameters.

The CLMS algorithm updates the filter coefficients via [4]

$$\mathbf{w}_n = \mathbf{P}[\mathbf{w}_{n-1} + \mu(y_n - \mathbf{w}_{n-1}^\top \mathbf{x}_n)\mathbf{x}_n] + \mathbf{q} \quad (3)$$

where

$$\mathbf{P} = \mathbf{I}_L - \mathbf{C}(\mathbf{C}^\top \mathbf{C})^{-1}\mathbf{C}^\top,$$

$$\mathbf{q} = \mathbf{C}(\mathbf{C}^\top \mathbf{C})^{-1}\mathbf{f},$$

$\mu$ is the step-size parameter, and $\mathbf{I}_L$ is the $L \times 1$ identity matrix.

## III. ANALYSIS

To make the analysis more tractable, let us use the following common assumptions [29], [30]:

*A1*: The input vectors of different time instants are independent zero-mean multivariate Gaussian and have a positive-definite covariance matrix $\mathbf{R} \in \mathbb{R}^{L \times L}$.

*A2*: The background noise is temporally-independent zero-mean Gaussian with variance $\eta \in \mathbb{R}_{\geq 0}$. It is also independent of the input data.

Under *A1* and *A2*, the optimal filter coefficient vector is given by [1]

$$\mathbf{g} = \mathbf{h} + \mathbf{R}^{-1}\mathbf{C}(\mathbf{C}^\top \mathbf{R}^{-1}\mathbf{C})^{-1}(\mathbf{f} - \mathbf{C}^\top \mathbf{h}).$$

Define the *deviation* vector by

$$\mathbf{d}_n = \mathbf{w}_n - \mathbf{g}.$$

Substituting (1) into (3), subtracting $\mathbf{g}$ from both sides of (3), and using

$$\mathbf{q} + \mathbf{P}\mathbf{g} - \mathbf{g} = \mathbf{0}_L$$

gives

$$\mathbf{d}_n = \mathbf{P}(\mathbf{I}_L - \mu \mathbf{x}_n \mathbf{x}_n^\top)\mathbf{d}_{n-1} + \mu \mathbf{P}\mathbf{x}_n \mathbf{x}_n^\top \mathbf{e} + \mu v_n \mathbf{P}\mathbf{x}_n. \quad (4)$$

Here, $\mathbf{0}_L$ is the $L \times 1$ zero vector and we define

$$\mathbf{e} = \mathbf{h} - \mathbf{g}.$$

The matrix $\mathbf{P}$ is idempotent, i.e., we have $\mathbf{P}^2 = \mathbf{P}$, which can be easily verified. Therefore, pre-multiplying both sides of (4) by $\mathbf{P}$ reveals that

$$\mathbf{P}\mathbf{d}_n = \mathbf{d}_n \quad \forall n.$$

Consequently, we can rewrite (4) as

$$\mathbf{d}_n = (\mathbf{I}_L - \mu \mathbf{P}\mathbf{x}_n \mathbf{x}_n^\top \mathbf{P})\mathbf{d}_{n-1} + \mu \mathbf{P}\mathbf{x}_n \mathbf{x}_n^\top \mathbf{e} + \mu v_n \mathbf{P}\mathbf{x}_n. \quad (5)$$

### A. Mean-square stability

Denote the Euclidean norm of a vector $\mathbf{b} \in \mathbb{R}^{L \times 1}$ by $\|\mathbf{b}\|$ and define its weighted Euclidean norm with a weighting matrix $\mathbf{A} \in \mathbb{R}^{L \times L}$ as

$$\|\mathbf{b}\|_\mathbf{A} = \|\mathbf{b}\|_{\text{vec}\{\mathbf{A}\}} = \sqrt{\mathbf{b}^\top \mathbf{A} \mathbf{b}}$$

where $\text{vec}\{\cdot\}$ is the vectorization operator that stacks the columns of its matrix argument on top of each other.

Bearing in mind *A1* and *A2*, calculating the expected value of the squared Euclidean norm of both sides of (5) yields the following variance relation:

$$E[\|\mathbf{d}_n\|^2] = E[\|\mathbf{d}_{n-1}\|^2_\mathbf{M}] + \mu^2 \mathbf{e}^\top E[\mathbf{x}_n \mathbf{x}_n^\top \mathbf{P} \mathbf{x}_n \mathbf{x}_n^\top]\mathbf{e} \\ + \mu^2 E[v_n^2 \mathbf{x}_n^\top \mathbf{P} \mathbf{x}_n] \quad (6)$$

where

$$\mathbf{M} = E[(\mathbf{I}_L - \mu \mathbf{P}\mathbf{x}_n \mathbf{x}_n^\top \mathbf{P})(\mathbf{I}_L - \mu \mathbf{P}\mathbf{x}_n \mathbf{x}_n^\top \mathbf{P})] \\ = \mathbf{I}_L - 2\mu \mathbf{Z} + \mu^2 \mathbf{P} E[\mathbf{x}_n \mathbf{x}_n^\top \mathbf{P} \mathbf{x}_n \mathbf{x}_n^\top]\mathbf{P} \quad (7)$$

and

$$\mathbf{Z} = \mathbf{P}\mathbf{R}\mathbf{P}.$$

Using the Isserlis' theorem [31] and *A1*, we get

$$E[\mathbf{x}_n \mathbf{x}_n^\top \mathbf{P} \mathbf{x}_n \mathbf{x}_n^\top] = E[\mathbf{x}_n^\top \mathbf{P} \mathbf{x}_n]E[\mathbf{x}_n \mathbf{x}_n^\top] \\ + 2E[\mathbf{x}_n \mathbf{x}_n^\top]\mathbf{P}E[\mathbf{x}_n \mathbf{x}_n^\top] \quad (8) \\ = \text{tr}\{\mathbf{Z}\}\mathbf{R} + 2\mathbf{R}\mathbf{P}\mathbf{R}.$$

Moreover, due to *A1* and *A2*, we have

$$E[v_n^2 \mathbf{x}_n^\top \mathbf{P} \mathbf{x}_n] = \eta \text{tr}\{\mathbf{Z}\} \quad (9)$$

and

$$\mathbf{P}\mathbf{R}\mathbf{e} = \mathbf{0}_L. \quad (10)$$

Substituting (8)-(10) into (6) and (7) gives

$$E[\|\mathbf{d}_n\|^2] = E[\|\mathbf{d}_{n-1}\|^2_\mathbf{M}] + \mu^2 \text{tr}\{\mathbf{Z}\}(\mathbf{e}^\top \mathbf{R}\mathbf{e} + \eta) \quad (11)$$

and

$$\mathbf{M} = \mathbf{I}_L - 2\mu\mathbf{Z} + \mu^2 \text{tr}\{\mathbf{Z}\}\mathbf{Z} + 2\mu^2 \mathbf{Z}^2.$$

The matrix $\mathbf{Z}$ has $K$ zero and $L-K$ nonzero eigenvalues, $\lambda_i, i=1,\ldots,L-K$ [4]. Subsequently, $\mathbf{M}$ has $K$ unit and $L-K$ non-unit eigenvalues, $\rho_i, i=1,\ldots,L-K$. The recursion of (11) is stable and convergent if

$$\rho_i < 1, i=1,\ldots,L-K$$

or equivalently

$$1 - 2\mu\lambda_i + \mu^2 \text{tr}\{\mathbf{Z}\}\lambda_i + 2\mu^2 \lambda_i^2 < 1, \quad i=1,\ldots,L-K. \quad (12)$$

To satisfy (12), it is enough to choose the step-size such that

$$0 < \mu < \frac{2}{2\lambda_{\max} + \text{tr}\{\mathbf{Z}\}} \quad (13)$$

where $\lambda_{\max}$ is the largest eigenvalue of $\mathbf{Z}$. Note that the mean-square stability upper-bound for $\mu$ in (13) is the same as the one given in [4] and [7] although our analytical approach is different from those of [4] and [7].

### B. Instantaneous MSD

Take $\mathbf{S} \in \mathbb{R}^{L \times L}$ as an arbitrary symmetric nonnegative-definite matrix. Applying the expectation operator to the squared-weighted Euclidean norm of both sides in (4) while considering *A1* and *A2* leads to the following *weighted variance relation*:

$$E[\|\mathbf{d}_n\|^2_\mathbf{S}] = E[\|\mathbf{d}_{n-1}\|^2_\mathbf{T}] + \mu^2 \mathbf{e}^\top E[\mathbf{x}_n \mathbf{x}_n^\top \mathbf{P}\mathbf{S}\mathbf{P}\mathbf{x}_n \mathbf{x}_n^\top]\mathbf{e} + \mu^2 E[v_n^2 \mathbf{x}_n^\top \mathbf{P}\mathbf{S}\mathbf{P}\mathbf{x}_n] \quad (14)$$

where

$$\mathbf{T} = E[(\mathbf{I}_L - \mu\mathbf{x}_n\mathbf{x}_n^\top)\mathbf{P}\mathbf{S}\mathbf{P}(\mathbf{I}_L - \mu\mathbf{x}_n\mathbf{x}_n^\top)] \\ = \mathbf{P}\mathbf{S}\mathbf{P} - \mu\mathbf{R}\mathbf{P}\mathbf{S}\mathbf{P} - \mu\mathbf{P}\mathbf{S}\mathbf{P}\mathbf{R} + E[\mathbf{x}_n\mathbf{x}_n^\top \mathbf{P}\mathbf{S}\mathbf{P}\mathbf{x}_n \mathbf{x}_n^\top] \quad (15)$$

In the same vein as (8) and (9), we have

$$E[\mathbf{x}_n\mathbf{x}_n^\top \mathbf{P}\mathbf{S}\mathbf{P}\mathbf{x}_n\mathbf{x}_n^\top] = \text{tr}\{\mathbf{S}\mathbf{Z}\}\mathbf{R} + 2\mathbf{R}\mathbf{P}\mathbf{S}\mathbf{P}\mathbf{R} \quad (16)$$

and

$$E[v_n^2 \mathbf{x}_n^\top \mathbf{P}\mathbf{S}\mathbf{P}\mathbf{x}_n] = \eta \text{tr}\{\mathbf{S}\mathbf{Z}\}. \quad (17)$$

Using (16), (15) can be written as

$$\mathbf{T} = (\mathbf{I}_L - \mu\mathbf{R})\mathbf{P}\mathbf{S}\mathbf{P}(\mathbf{I}_L - \mu\mathbf{R}) + \mu^2 \text{tr}\{\mathbf{S}\mathbf{Z}\}\mathbf{R} + \mu^2 \mathbf{R}\mathbf{P}\mathbf{S}\mathbf{P}\mathbf{R} \quad (18)$$

Applying the vectorization operator to (18) together with using the properties [32]

$$\text{vec}\{\mathbf{A}\mathbf{B}\mathbf{C}\} = (\mathbf{C}^\top \otimes \mathbf{A})\text{vec}\{\mathbf{B}\}$$

and

$$\text{tr}\{\mathbf{A}^\top \mathbf{B}\} = \text{vec}^\top\{\mathbf{B}\}\text{vec}\{\mathbf{A}\} \quad (19)$$

yields

$$\text{vec}\{\mathbf{T}\} = \mathbf{F}\mathbf{s} \quad (20)$$

where

$$\mathbf{F} = [(\mathbf{I}_L - \mu\mathbf{R})\mathbf{P} \otimes (\mathbf{I}_L - \mu\mathbf{R})\mathbf{P}] + \mu^2 \text{vec}\{\mathbf{R}\}\text{vec}^\top\{\mathbf{Z}\} \\ + \mu^2 (\mathbf{R}\mathbf{P} \otimes \mathbf{R}\mathbf{P}),$$

$$\mathbf{s} = \text{vec}\{\mathbf{S}\},$$

and $\otimes$ denotes the Kronecker product. Substituting (16), (17), and (20) into (14) together with using (10) and (19) gives

$$E[\|\mathbf{d}_n\|^2_\mathbf{s}] = E[\|\mathbf{d}_{n-1}\|^2_\mathbf{Fs}] + \mu^2(\mathbf{e}^\top \mathbf{R}\mathbf{e} + \eta)\text{vec}^\top\{\mathbf{Z}\}\mathbf{s}. \quad (21)$$

By making appropriate choices of $\mathbf{s}$ in (21), for any time instant $n$, we can write

$$E\left[\|\mathbf{d}_i\|^2_{\mathbf{F}^{n-i}\mathbf{j}}\right] = E\left[\|\mathbf{d}_{i-1}\|^2_{\mathbf{F}^{n-i+1}\mathbf{j}}\right] \\ + \mu^2(\mathbf{e}^\top \mathbf{R}\mathbf{e} + \eta)\text{vec}^\top\{\mathbf{Z}\}\mathbf{F}^{n-i}\mathbf{j}, \quad 1 \leq i < n \quad (22)$$

where we define

$$\mathbf{j} = \text{vec}\{\mathbf{I}_L\}.$$

Summation of both sides in (22) for $i=1,\ldots,n$ gives

$$E[\|\mathbf{d}_n\|^2] = \|\mathbf{d}_0\|^2_{\mathbf{F}^n\mathbf{j}} + \mu^2(\mathbf{e}^\top \mathbf{R}\mathbf{e} + \eta)\text{vec}^\top\{\mathbf{Z}\}\sum_{i=0}^{n-1}\mathbf{F}^i\mathbf{j}. \quad (23)$$

Similarly, we can show that

$$E[\|\mathbf{d}_{n-1}\|^2] = \|\mathbf{d}_0\|^2_{\mathbf{F}^{n-1}\mathbf{j}} \\ + \mu^2(\mathbf{e}^\top \mathbf{R}\mathbf{e} + \eta)\text{vec}^\top\{\mathbf{Z}\}\sum_{i=0}^{n-2}\mathbf{F}^i\mathbf{j}. \quad (24)$$

Subtraction of (24) from (23) results in the time-evolution recursion of the instantaneous MSD as

$$E[\|\mathbf{d}_n\|^2] = E[\|\mathbf{d}_{n-1}\|^2] - \|\mathbf{d}_0\|^2_{\mathbf{F}^{n-1}(\mathbf{I}_{L^2}-\mathbf{F})\mathbf{j}} \\ + \mu^2(\mathbf{e}^\top \mathbf{R}\mathbf{e} + \eta)\text{vec}^\top\{\mathbf{Z}\}\mathbf{F}^{n-1}\mathbf{j}.$$

### C. Steady-state MSD

Provided that (13) is fulfilled, the CLMS algorithm converges in the mean-square sense. Thus, at the steady state, i.e., when $n \to \infty$, (21) turns into

$$\lim_{n\to\infty} E[\|\mathbf{d}_n\|^2_\mathbf{s}] = \lim_{n\to\infty} E[\|\mathbf{d}_n\|^2_\mathbf{Fs}] + \mu^2(\mathbf{e}^\top \mathbf{R}\mathbf{e} + \eta)\text{vec}^\top\{\mathbf{Z}\}\mathbf{s}$$

or

$$\lim_{n\to\infty} E\left[\|\mathbf{d}_n\|^2_{(\mathbf{I}_{L^2}-\mathbf{F})\mathbf{s}}\right] = \mu^2(\mathbf{e}^\top\mathbf{R}\mathbf{e}+\eta)\text{vec}^\top\{\mathbf{Z}\}\mathbf{s}. \quad (25)$$

Choosing

$$\mathbf{s} = (\mathbf{I}_{L^2} - \mathbf{F})^{-1}\text{vec}\{\mathbf{I}_L\}$$

and substituting it into (25) results in the steady-state MSD of the CLMS algorithm:

$$\lim_{n\to\infty} E[\|\mathbf{d}_n\|^2] = \mu^2(\mathbf{e}^\top\mathbf{R}\mathbf{e}+\eta)\text{vec}^\top\{\mathbf{Z}\}(\mathbf{I}_{L^2}-\mathbf{F})^{-1}\text{vec}\{\mathbf{I}_L\}.$$

*D. Steady-state misadjustment*

The steady-state misadjustment of the CLMS algorithm is defined as [33]

$$\zeta = \lim_{n\to\infty} \frac{E[(y_n - \mathbf{w}_{n-1}^\top\mathbf{x}_n)^2] - E[(y_n - \mathbf{g}^\top\mathbf{x}_n)^2]}{E[(y_n - \mathbf{g}^\top\mathbf{x}_n)^2]}$$

and can also be expressed as

$$\zeta = \frac{\lim_{n\to\infty} E[\|\mathbf{d}_{n-1}\|^2_\mathbf{R}]}{\mathbf{e}^\top\mathbf{R}\mathbf{e}+\eta}. \quad (26)$$

Setting

$$\mathbf{s} = (\mathbf{I}_{L^2} - \mathbf{F})^{-1}\text{vec}\{\mathbf{R}\}$$

in (25) gives

$$\lim_{n\to\infty} E[\|\mathbf{d}_n\|^2_\mathbf{R}] = \mu^2(\mathbf{e}^\top\mathbf{R}\mathbf{e}+\eta)\text{vec}^\top\{\mathbf{Z}\}(\mathbf{I}_{L^2}-\mathbf{F})^{-1}\text{vec}\{\mathbf{R}\}. \quad (27)$$

Using (27) in (26), we get

$$\zeta = \mu^2\text{vec}^\top\{\mathbf{Z}\}(\mathbf{I}_{L^2}-\mathbf{F})^{-1}\text{vec}\{\mathbf{R}\}. \quad (28)$$

Note that although (28) is seemingly different from the expression derived in [6] and [7] for the steady-state misadjustment, i.e.,

$$\zeta = \frac{\sum_{i=1}^{L-K} \frac{\mu\lambda_i}{1-\mu\lambda_i}}{2 - \sum_{i=1}^{L-K} \frac{\mu\lambda_i}{1-\mu\lambda_i}}, \quad (29)$$

it can be verified that (28) and (29) are in fact identical.

SIMULATIONS

Consider a constrained system identification problem where the underlying linear system is of order $L = 7$ and there exist $K = (L-1)/2$ linear equality constraints. We set the system parameter vector, $\mathbf{h}$, the constraint parameters, $\mathbf{C}$ and $\mathbf{f}$, and the input covariance matrix, $\mathbf{R}$, arbitrarily. However, we ensure that $\mathbf{h}$ has unit energy, $\mathbf{C}$ is full-rank, and $\mathbf{R}$ is symmetric positive-definite with $\text{tr}\{\mathbf{R}\} = L$. The input vectors are zero-mean multivariate Gaussian. The noise is also zero-mean Gaussian. We attain the experimental results by averaging over $10^4$ independent runs and, when applicable, over $10^3$ steady-state values.

In Fig. 1, we depict the theoretical and experimental MSD-versus-time curves of the CLMS algorithm for different value of the step-size when the noise variance is $\eta = 10^{-2}$.

In Fig. 2, we plot the theoretical and experimental steady-state MSDs of the CLMS algorithm as a function of the noise variance for different values of the step-size.

In Fig. 3, we compare the theoretical and experimental values of the steady-state misadjustment for different step-sizes. We include both (28) and (29) as well as the lower and upper bounds given in [4], i.e.,

$$\zeta_{\min} = \frac{\mu\,\text{tr}\{\mathbf{Z}\}}{2 - \mu(\text{tr}\{\mathbf{Z}\} + 2\lambda_{\min})}$$

and

$$\zeta_{\max} = \frac{\mu\,\text{tr}\{\mathbf{Z}\}}{2 - \mu(\text{tr}\{\mathbf{Z}\} + 2\lambda_{\max})}$$

where $\lambda_{\min}$ is the smallest nonzero eigenvalues of $\mathbf{Z}$. Fig. 3 shows that (28) and (29) are equivalent.

Figs. 1-3 illustrate an excellent match between theory and experiment, verifying the analytical performance results developed in this paper.

IV. CONCLUSION

We studied the mean-square performance of the constrained least mean-square algorithm and derived theoretical expressions for its transient and steady-state mean-square deviation. Through simulation examples, we substantiated that the resultant expressions are accurate for a wide range of values of the noise variance and step-size parameter. The presented theoretical formula can help designers predict the steady-state performance of the CLMS algorithm and tune its step-size to attain a desired performance in any given scenario without resorting to Monte Carlo simulations.

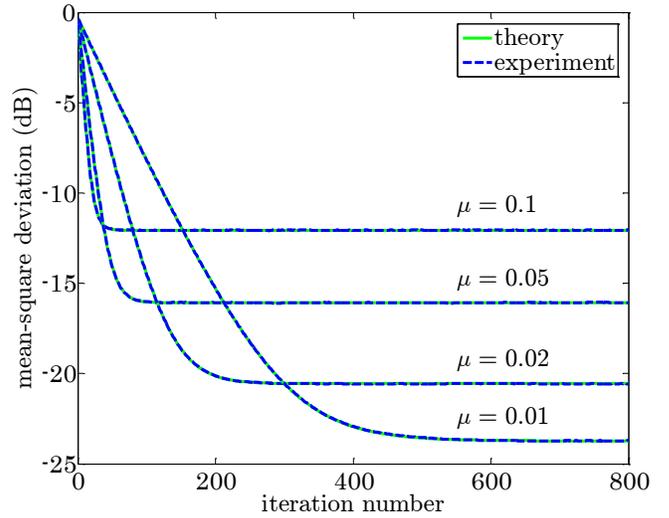

Fig. 1. MSD of the CLMS algorithm versus the iteration number for different values of the step-size when $\eta = 10^{-2}$.

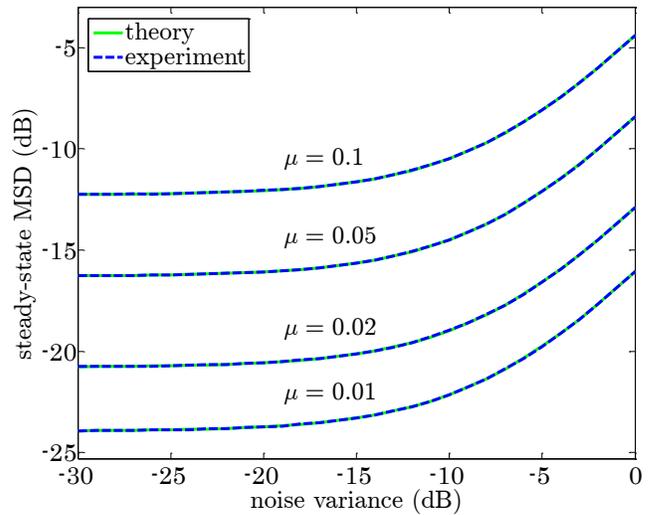

Fig. 2. Steady-state MSD of the CLMS algorithm versus the noise variance for different values of the step-size.

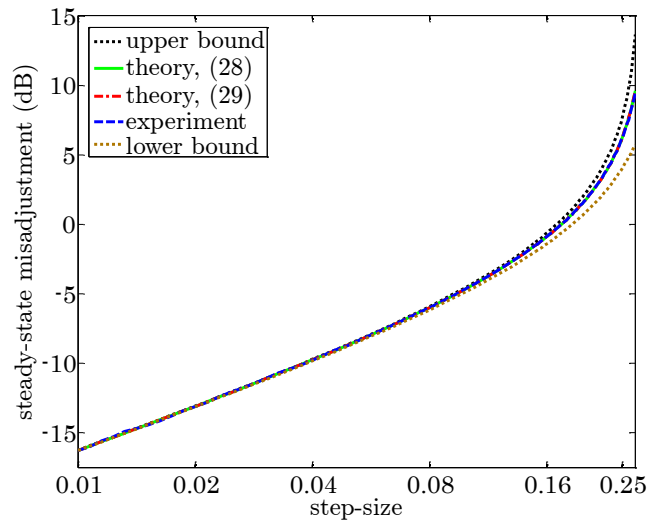

Fig. 3. Steady-state misadjustment of the CLMS algorithm versus the step-size.